# Data mining the MNC like internal co-opetition duality in a university context.

## Abstract

The goal of the paper is to quantify the simultaneous competition and cooperation that takes place in organizations. As the concepts seem to be dichotomous opposites at first, the term internal co-opetition duality is put forth. Parallels are drawn between co-opetitive processes in big multinational corporations (MNCs) and these taking place in universities, also the structural solutions used in both are analyzed.

Data mining is used while looking at how specializations inside the university are in competition for "better" students. We look at the profiles that students have and find natural divisions between the specializations, by applying graph theory and modularity algorithms for community detection. The competitive position of the specializations is evident in the average grades of the detected communities. The ratio of intercommunity ties to intracommunity ties (conductance) quantifies the cooperative stance, though, as students with similar profiles express linkages in the curricula; and the choices regarding career development undertaken become evident. Managerial implications discussed include the imperative for actively managing and financially rewarding the outcomes of the co-opetitive duality.

**Introduction**

A decade and a half ago, the term co-opetition was coined (Brandenburger and Nalebuff, 1998), indicating the simultaneous competition and cooperation that takes place on the markets. By now there have been calls for (Walley, 2007) and some efforts of studying it on other levels - e.g. society, intrafirm, personal (Ritala *et al.*, 2009; Walker *et al.,* 2004). There is, though, some confusion as to how the simultaneous competition and cooperation process can actually pan out.

Our research considers the co-opetitive state of affairs inside universities. For one facet of our research, we are convinced that a wide range of business theories, especially those dealing with huge multinational corporations (MNCs), can be used to inform us about processes taking place in educational institutions. Such companies form network organizations with natural national boundaries and strong implications for competition-cooperation.

Our goal is to empirically quantify the co-opetition taking place - this way we can create means for actively managing the co-opetitive process. We will apply network science by analyzing the similarities of student profiles.

This paper is organized in the following way: Part 1 analyzes the relationship that competition and cooperation have; Part 2 speaks of different conceptualizations of the process on societal level; Part 3 brings parallels between the situation in MNCs and universities for the process; Part 4 analyzes the structural solutions used, that are relevant for our analysis; finally Part 5 puts forth the empirical investigation of the process in Tallinn University of Technology and the managerial implications.

## 1. The co-opetition duality

In a dichotomy, a whole is split into two non-overlapping parts. In a classic article on "concept misformation", Sartori (1970)* argues that concept formation is inherently based on classification and that dichotomies are exclusively fundamental to reasoning about concepts. However, a large body of research in linguistics, cognitive psychology and cognitive science is supporting a more multifaceted view of human cognition, according to which the remarkable capacity of the mind to conceptualize different modes of gradation and different forms of the partial occurrence of phenomena, is equally important (Collier and Adcock, 1999). In linguistics, for example, cline is a scale of continuous gradation. Therefore, as both verbs *see* and *kill* are transitive (as opposed to intransitive), *see* is described as having lower transitivity.

When dichotomous concepts at the ends of the continuum are each other's opposites, a paradox is formed (Poole and Van de Ven, 1989), marking the seeming impossibility of seamlessly integrating the two. An example here would be the objects of our research, the processes of competition and cooperation, in case of which it is difficult to conceptualize a 70% competitive and 30% cooperative arrangement. The paradox of simultaneous competition and cooperation is in business literature termed co-opetition (Branderburger and Nalebuff, 1998). But this is something that we see taking place inside organizations, just as between them, between individuals in a society, and also elsewhere in the nature - for example, for bacteria (Griffin *et al.,* 2004).

Chen (2008) integrates the Western term paradox and the Eastern "middle-way" thinking into a concept transparadox, in case of co-opetition using three levels - independent opposites, interrelated opposites and interdependent opposites. An example of independent opposites would be the strict choice between competition and cooperation that oligopoly market theory presents for neoclassical economics (Scherer and Ross, 1990*). For interrelated opposites, an example would be a US car company offering a 1000$ rebate on car parts, redeemable in an outlet of any market participant (Chen, 2008). We are the most concerned with the third option - the interdependent opposites. It is described as is the relationship between light and dark - as something, wherein one is defined through the other, wherein the two do not have an independent meaning.

We borrow the term "duality" from Chen (2008) to mark what is meant by the third kind of simultaneous competition and cooperation, that also takes place within organizations. Duality is widely used in mathematics, for instance, in operations research, where a problem and its dual are solved in an interdependent fashion and the values of their maximum/minimum are the same (eg. Bazaraa *et al.,* 1990*).

Thus, we study the internal co-opetition duality. As co-opetition has been in research focus for more than a decade and a half by now (eg. Peng and

Bourne, 2009; Ritala, 2011; see Peng *et al.,* 2011 for another application of the Eastern way of thinking), a need for studying internal co-opetition has only been stated recently (Walley, 2007). There have only been some studies, e.g. Ritala *et al.* (2009) considering the link of internal co-opetition with knowledge transfer and innovation, as well as a narrowly published earlier effort (Ubi, 2003).

## 2. Co-opetition duality in society

There are a number of bodies of work, which consider competition and cooperation as dichotomies, as "independent opposites", in situations where they actually are not, and where thus the recognition of the duality of co-opetition would be of benefit.

We start on the societal level, with a strand of research advocating the "feminist position" (e.g. Harding, 2004), which while being informed by neuropsychiatrist research on the difference between women and men (e.g. Brizendine, 2006), states that the male economists have asked questions and drawn conclusions only in a certain way and have not incorporated the "women's way of knowing" (Ferber and Nelson, 1993). It posits that the masculine scientific mainstream has overly emphasized competition, which has its negative connotations, whereas cooperation should be at helm instead.

The fact that this strand of research draws parallels with the masculinity and femininity dimension in a society is noteworthy. Masculinity-femininity dimension is, according to the studies assessing national cultures (Hofstede, 2005), the only cultural dimension that differentiates countries even after we take into account its wealth - thus coloring our understanding of co-opetition duality in international organizations and organizations internationally.

This paper sides with the critics of the feminist position (c.f. Walker *et al.,* 2004), who speak of simultaneous competition and cooperation - on the societal level. The basic way of reasoning goes, that in order to compete, there has to be an agreement upon the rules of the process - cooperation - in the first place. Also, as made evident by the division of labor, rational humans learned that cooperative action is more efficient than isolated action. As has become evident throughout the 20th century, cooperation without competition will lead to stagnation, as competition is a discovery procedure, providing us with the signals from the market. Neither component of this duality, competition nor cooperation, can be stated to be the sole "rectified" final goal for humans - as in the case of cooperation, our final goal would stem from the social nature of humans, and in the case of competition, it would be to solely increase the efficiency and material progress.

## 3. Co-opetition duality in organizations - parallels between multinational corporations and universities

The basic premise of our research into the co-opetition duality, is the comparison of the state of affairs in educational organizations with those prevailing in commercial organizations - more specifically the multinational corporations (MNCs).

In MNCs the basic units operate in different countries and are called subsidiaries. In the university under consideration - Tallinn University of Technology (TUT) - the basic units of organization are the departments, and there is as strong of a separation between these, as there is between the subsidiaries of an MNC. If a faculty member was to move from one department to another, for the management, it would be as if he had left the organization altogether.

Humes (1993) speaks of managers on a fast track and, thus, of creating carriers of MNC corporate culture. MNCs are essentially three dimensional

(Ubi, 2003), as they have people working at different functions of the company (R&D, marketing, production, etc.); also working on different products/in different product divisions; and finally in different countries. Corporate culture is an important tool for MNCs (Hedlund, 1986). According to Humes (1993) different bonds will have to be broken while creating its carriers. For better carrying of corporate culture an employee would have to be transferred between different functions (professional alignment broken); would have to change its position on product dimension; would have to work internationally. As was mentioned, transfers between departments of TUT are rare, and also, there is less of a chance for it, as often the content of the work differs fundamentally. This forms a reason for there to be less cooperation, than in MNCs - though the organization only has one basic separating dimension, departmental, there are no classically footloose employees, as in the case for managers on the fast track for MNCs.

On the other hand, as each MNC subsidiary is embedded in its local national culture, alignment with it is an additional factor influencing the co-opetition duality, as employees, who are not expatriates, identify mainly with their home country (subsidiary). Just as the ordinary faculty members in a university department do.

Let us now consider some structural solutions, that MNCs use for dealing with the simultaneous competition and cooperation. MNC subsidiaries may have developed into World Product Mandates (WPM) (Roth and Morrison 1992). This means that it has acquired control of the full value-added scope (logistics, R&D, production and marketing) of a specific product or product line, with the responsibility of producing for the world market and controlling the entire value chain. MNC subsidiaries may, on the other hand, also, be so successful at creating certain product divisions, say, marketing campaigns, that they are formally acknowledged as a marketing Center-of-Excellence for the Western hemisphere (COE) (Moore and Birkinshaw 1998). Figure 1 depicts these two structural embodiments (in red and blue) on a subset of a three dimensional map of an MNC.

**Figure 1. A divergent possibly overlapping structural map of an MNC.**

We can say that WPM is something, that manifests competition within an MNC. It is the result of development of a subsidiary (Kirstuks, 1999), whereby its capabilities are enhanced. This is best done by taking entrepreneurial action (e.g. Birkinshaw, 2000). For that internal corporate venturing may be used - the creation of divisions with a specific mission for innovation (Burgelman, 1983) -, or more relevantly for the situation within universities, by "intrapreneurship" (Ghoshal and Bartlett, 1997), wherein all members of an organization are expected to act on emerging opportunities. According to Birkinshaw (1997), entrepreneurial action is taken in the form of subsidiary initiatives, which can either be directed towards the external marketplace or internally. External initiatives would manifest development of department's capabilities, in the case of a university, whereas internal initiatives - winning something, like better performing students, from a competing department - are direct means of internal competition.

It has been widely documented, that MNC subsidiaries engage in internal competition (e.g. Krajewski *et al.,* 1994; Galunic and Eisenhardt, 1996), by out-competing a sister unit for an activity, already performed or upcoming. When the need for internal competition is deemed necessary to be emphasized, terms such as MNCs operating on internal markets are used (c.f. Birkinshaw, 2000; but also similarly Buckley and Casson, 1998). Increased competition is said to be evident from broad use of internal benchmarking and performance league tables, or operations of internal investment agencies.

After out-competing a sister subsidiary for the WPM production of its former "charter", it is natural that a less than friendly atmosphere might arise (c.f e.g. Birkinshaw,1995).

In a university, another form of internal competition takes place in the budgeting process (c.f. in business organizations Walley, 2007), wherein the amount of money that the state allocates by paying for the tuition of the better performing students is planned for.

At the same time, it is important to recognize the need for cooperation inside a business organization like an MNC - as all subsidiaries are still part of the same company. In management literature there is often a great emphasis that is being put on achieving widespread sharing and trust. When in the process of developing subsidiary capabilities leading edge solutions arise, it becomes a natural objective to disseminate these throughout the corporation (c.f. Andersson and Holmström, 2000).

In the case of MNCs' literature on knowledge transfer and organizational learning, the managers on a fast track are once again mentioned, just as well as development of leading edge IT solutions. Organizational culture needed for such process is characterized as open, based on fairness and shared values.

In service industries consulting companies, such as McKinsey, are a fine examples; while in the case of industrial firms IKEA stands out (Ghoshal and Bartlett, 1997; also see Heldlund and Ridderstrale, 1995 on projects of international cooperation).

Structurally, we will define COE as the manifestation of such best practice transferal, another example being 3M Sweden, which had leading capabilities in customer focused marketing and key account management. It was recognized for them and actively helped out other units (Birkinshaw, 2000), being an exemplary knowledge disseminator in an MNC.

In a university such cooperation is evident in the case of doctoral schools, that are jointly undertaken between the departments operating in the same field. In the case of interdisciplinary theses departments take part in joint supervisions, there can be as many as three supervisors, with industry representatives participating. There are also horizontal committees that discuss topics like curriculum development, student dropout reduction, industry cooperation, internationalization, etc.

The curricula in TUT must contain interlinked parts. There is a minimum number of credits, that have to be taken from other departments and faculties, thus creating natural linkages between the organizational units. The linkages are not only a manifestation of cooperation, though. They bring with it the monetary reward, as all the declarations have to be paid for - there are sometimes competing departments, who could provide the same courses.

Let us state, for our co-opetition duality, that world product mandates and centers-of-excellence are symbols of its interdependent opposites - that gaining a mandate (the right to be the sole proprietor of certain activities) and being a center (to be recognized as an asset valuable enough to be explicitly distributed) are goals toward which the organizational unit will strive simultaneously.

As we see, sometimes it is the competition that is being emphasized and sometimes cooperation. An example of the latter is the work of Eisenhardt and Galunic (2000). This article defines MNCs as "coevolving systems" and certainly emphasizes the fact that everyone is in the same boat therein.

Still, the two discussions are actually not always isolated. The internal market perspective (Birkinshaw, 2000) has also discussed the phenomenon of knowledge transfer. Corporations are said to have an internal market of capabilities, one that operates without competition and without fees charged for servicing. It is said to be facilitated by strong corporate culture and incentive systems that reward propensity to cooperate. HP and Ericsson are brought as examples of companies, that do a feasible job at sharing at the same time with competing.

# 4. Internal organization for co-opetition - multinational corporations and universities

A central point while considering, how MNCs try to find a balance between competition and cooperation, is that they should be considered as multicentered/networked, with less emphasis on the overarching hierarchy that might also have been defined. The same is true for universities, with its departmentally oriented structure. In the case of MNCs, the term used is heterarchical (Hedlund, 1986; Hedlund and Rolander, 1990; Hedlund, 1993), which points to organizations with a partially overlapping and divergent structural map of Figure 1, but is based on a change of perspectives in a wide range of sciences. The main idea being that the reality is actually organized non-hierarchically and that we are only accustomed to working with it through hierarchies. One good example from complex embodiments is the highest level in evolution - the brain -, the functioning of which we cannot completely explain, but which surely is a non-hierarchical system.

Important contributions are also made by the Uppsala school (e.g. Forsgren and Pedersen, 1998; Forsgren and Pahlberg, 1992), which often does not make qualitative differentiation between cases wherein an organizational unit is involved in transactions with sister units and cases wherein transactions take place with external parties. Some business network articles have also modeled the positions of subsidiaries within MNCs in terms of influence (Forsgren and Pahlberg, 1992; also Ghoshal and Bartlett, 1990) - something that we also aspire to do. Another point propounded by the Uppsala school, is that there also exists a continuum, when considering, whether a transaction takes place on the market - at *arm's length* -, or through strict *hierarchical fiat*. That is, in business networks, we should not consider all the business partners as strictly separable from their relationships. As the arm's length transaction would imply competition for resources and hierarchical fiat cooperation thereupon, we find the venue of research to be related with ours. Our empirical analysis will view the university as a network of units, one with a heterarchical structure. Startup companies originating from the university exemplify the fuzzy boundaries that the university has with the industry.

Another point of interest, are the works of Burgelman (1983, 1994) that consider autonomous strategic decisions in big corporations. These works show how multiple layers of management are actually involved in taking strategic decisions. This way the managers, who are at first to be considered lower at the organizational drawing board, can through their actions be setting strategic directions of the company. An example, in this case being Intel, in which the decisions of middle level managers decided the transferal of a memory company into a processor company (Burgelman, 1994). This set of articles also implies a multi-centered view of an organization, but is also valid in the context of universities, in the sense that there is usually an ample space for autonomous strategic initiatives in academia - be it for industry cooperation or research direction choice.

There is also another body of work that can be drawn in parallel with the discussion on academic freedom - namely research on "subsidiary slack". Poynter and White (1985) discuss organizational slack that develops in subsidiaries, and the ways of its dissipation. Slack is defined as the excess of total human resources after a proper amount has been allocated for the current strategy. The work shows that subsidiaries have a natural tendency of generating slack and that it can be used for undertaking new value-added activities. As with academic freedom, if initiatives arising from slack are not tolerated, disharmonies arise.

Be it for autonomous initiatives or slack, the resulting external initiatives are important for the development of the university - the fruits of which need to be shared cooperatively. Internal initiatives on the other hand enhance the position of the focal department and are directly related to the competitive placing within the university.

**5. Quantifying co-opetion in Tallinn University of Technology**

We are using data from Tallinn School of Economics and Business Administration (TSEBA) of TUT. It consists of 509 students, who have graduated the curriculum "Business" during the time span 1997-2010. We have information regarding the courses that the students have attended, the grades they received, as well as on the final specialization the student has chosen within the business school. Altogether, students have attended 759 courses. The courses were declared 21976 times. The four specializations of TSEBA are: finance, marketing, accounting and management.

**Figure 2. Binary matrix displaying the course selection of students.**

As the first step of analysis we construct a binary matrix (partially displayed on Figure 2), with rows representing the students and columns representing the courses attended. This matrix is sparse, with 5.66% of cells marked by 1, on the whole. The most frequently taken course has been attended 501 times, 463 courses have only been attended once; and 71 courses have been attended more than 20 times.

We will next construct a 509x509 distance matrix between all the students. As the first step we will use Hamming distances, the essence of which is finding the total number of different course selections by the two students, - thus arriving at a distance matrix. For the selection of the distance function, we also tried weighted Hamming distances (the "importance" of a course, the number of times attended, was used as a weight) but this led to significantly poorer community detection, as the course attendance distribution is strongly right skewed. We will next calculate the matrix of similarities by taking the reciprocal value of the distance. As the next step, we calculate the distribution function of the similarity matrix (on Figure 3), and find out that as the values range from 0..1, 95% of the values are below 0.11. Our final similarity matrix will be composed of those 5% of entries, that have value above 0.11 - thus it is once again sparse, with the same level of sparsity that the original course attendance matrix had.

**Figure 3. Distribution function of similarity matrix and cutoff value.**

**Figure 4. Detection of student specializations for competitive and cooperative stance.**

On Figure 4, we will use ForceAtlas2 algorithm (Jacomy *et al.*, 2011) in order to visualize the similarity matrix/adjacency matrix as a social network. This algorithm has a linear-linear model, with attraction and repulsion forces proportional to the distance between the nodes. The shape of its graph is between Frücterman&Reingold's layout and Noack's LinLog.

As the next step (also on Figure 4), we are going to detect and color the communities. In order to quantify the intercommunity ties, we need an algorithm, that does not only find (in sparse graphs) isolated communities, with different levels cliquishness (like Closed Sets algorithms – eg. Lohk *et al.,* 2010; see Hruschka, 2006**, for cliquishness), nor the isolated perfect clicks (like Formal Concept Analysis eg. Torim and Lindroos, 2008), but one that also allows for intercommunity ties. We have chosen the Modularity algorithm (Blondel *et al.*, 2008), which in its class is a comparatively fast performing greedy heuristic (cf. Fortunato, 2010). It builds hierarchical communities in two iterating passes, by joining the nodes and building a new structure thereupon. New nodes are being joined based on intracommunity linkages, linkages outside the community and the same two kinds of linkages for the focal node being joined.

As a result, specialization is correctly predicted for 65% of students (331 out of 509). These form the BLUE, YELLOW, GREEN and CYAN communities on the graph. 21% of students (108 out of 509) are joined in the RED community, which incorrectly consists of students of all specializations in approximately equal percentages. 13% of students (70 out of 509) are incorrectly not included in any community (these are the solitary dots on the "outside").

The communities follow the logical structure of the business school -

BLUE - marketing,

YELLOW - management,

GREEN - finance,

CYAN - accounting -

by the following token: management uses information from marketing and finance; finance is tied to management as its internal consumer, but also receives data from accounting; the distances showing the logical distances between the specializations.

As shown on Figure 4 all the correctly detected communities are split into two. The factors accounting for this division are a subject for a follow-up discriminatory analysis. Furthermore, the discriminatory factors accounting for the emergence of the RED group should also be studied. On another note, the two sides of the CYAN accounting group were also originally split into two, as we would have done by using the discriminant analysis.

Average grades of the groups (on 5 point scale), representing the competitive position of the department "mandated" with the specialization, are as follows:

BLUE - marketing (3.38),

YELLOW - management (3.15),

GREEN - finance (3.64),

CYAN - accounting (3.35).

As the strong intragroup linkages define the specializations and allow them to be detected, they express students, who have small Hamming distances and correspondingly similar student profiles. If the students, whom the specialization has been "mandated" to teach, have a high average grade, the specialization is doing competitively well, otherwise not.

The sparse similarity matrix, with 5% of its values nonzero, though, also displays linkages between the groups. This means, that there are students, whose profiles are similar, but whose specializations do not coincide. These depict the linkages in curricula, the fact that a student studying, for example, managerial finance, will concentrate both on managerial accounting and on corporate finance. Thus, this is a student, who will constitute a link between the accounting and the finance "centers" of the graph. As elaborated above, these linkages also display financial ties; and the fact that departments are required to tie together the curricula.

For the sparse similarity matrix, we consider the following ratio, as one expressing the cooperative stance that a specialization has:

$$Cooperation = \frac{\sum all\ linkages\ from\ the\ focal\ specialization\ leading\ to\ other\ groups}{\sum all\ intragroup\ linkages\ for\ the\ focal\ specialization}$$

or denoting differently in social network analysis terms as traditional conductance (Mislove *et al.*, 2007)

$$K = \frac{\varepsilon_{AB}}{\varepsilon_{AA}}$$

We choose traditional conductance over very similar ones like relative conductance (Mislove *et al.*, 2007) for the comfortableness of the interpretation of its results. If the ratio is high, the specialization is doing cooperatively well, otherwise not:

BLUE - marketing (12.78%),

YELLOW - management (33.91%),

GREEN - finance (32.85%),

CYAN - accounting (15.75%).

As can be summarized from the competitive and cooperative positions, finance is the specialization that does very well on both accounts. It has the best competitive position and almost the highest cooperation ratio as well. Managerial implications for university management concern the evolving dynamics of co-opetition duality. Clearly, the competitive stance is important. But so is the fact, that university curricula must consist of intertwined parts. The combination of the two must at least be recognized, but can also be financially rewarded. In the case only the competitive position would be rewarded, the university departments would ultimately diverge in their performance – with the strong only getting stronger. On the other hand it would not be correct to punish success, but instead also reward cooperation that emerges between the departments.

**Conclusion**

For the situation inside organizations, this paper disagrees with the viewpoint, whereby competition and cooperation form a dichotomous paradox, having either one or the other present. Instead we side with the call for using the Eastern "middle-way" thinking, which names this relationship a duality. In a duality the two interdependent sides cannot exist without, and are defined by, each other - like, for instance, light and dark. As one might intuitively vouch for there to be more competition on the markets and more cooperation on the intrafirm level, we set out to study the internal co-opetition duality.

While considering studies on the societal level, we see the same question of the nature of the co-opetitive relationship raised over again. We next compare the state of affairs in big multinational corporations with these prevailing in the universities.

We use the term World Product Mandate as an exemplifier of the competition that takes place inside MNCs. At the same time Center Of Excellence shows that a national subsidiary of an MNC has been effective in dispersing its unique capabilities. Thus, the difference between being **mandated** and being a **center** throws into relief the nature of the duality of co-opetition in an MNC. A mandate is achieved by either internal or external initiatives. In a university context the former would mean the development of departmental capabilities - by research or industry consultation - whereas the latter implies direct competition for better students. Competition also takes place in universities during the budgeting process. While both MNCs and universities have an innate tendency to act parocially, it might be easier to be cooperative for MNCs due to the use of managers on a fast track and greater internal mobility of workforce in general. At the same time studies of MNCs underline widespread sharing and trust that has to exist. One of the studies, that emphasizes cooperation, is considering organizations as coevolving systems. Studies of internal competition have also considered cooperative aspects. In universities, cooperation is evident from doctoral schools, joint supervisions, horizontal commitees and interlinked curricula. Interlinked curricula has cooperative, but also competitive aspects.

Organizationally the structure of MNCs has been described as heterarchical, networked and leaving space for autonomous initiatives. As for the boundaries of the organization, it is said to be somewhere between the *hierarchical fiat* and *arm's length transactions.* We find university structure to also be networked and with fuzzy boundaries. Works on MNCs have considered subsidiray slack, that is related to autonomous initiatives. Slack is a concept that is also relevant in the university context, as there has to be academic freedom in this institution.

In order to quantify the co-opetitive duality, we consider a binary sparse matrix of the 509 graduates of the business school of Tallinn University of Technology, and their 759 courses declared. We calculate the reciprocal of Hamming distances between the students and are only concerned with the top 5% of the ties between the students. We use ForceAtlas2 algorithm - one with linear-linear model - for calculating the layout of the adjacency matrix. We next use a fast two-stage iterative heuristic - Modularity algorithm - in order to detect interrelated communities in the social network. As there are four specializations in the business school - marketing, management, finance and accounting - 65% of the students are correctly detected by their specialization using the Modularity algorithm. Further discriminant analysis is warranted in order to delineate the characteristics of the undetected group, as well as the binary splits inside the specializations. The average grade of the detected group makes evident the competitive position of the department inside the business school. At the same time, the ratio of intercommunity ties to intracommunity ties (traditional conductance in a social network) quantifies the cooperative stance. According to both, the best performer is clearly the finance specialization. There is a need though, to actively manage the co-opetitive duality, taking into account both forces simultaneously. The managerial implication for driving the evolution of the dynamics of the co-opetition duality is to neither only let strong get stronger nor punish good performance, but to take cooperation into account together with the competitive stance.

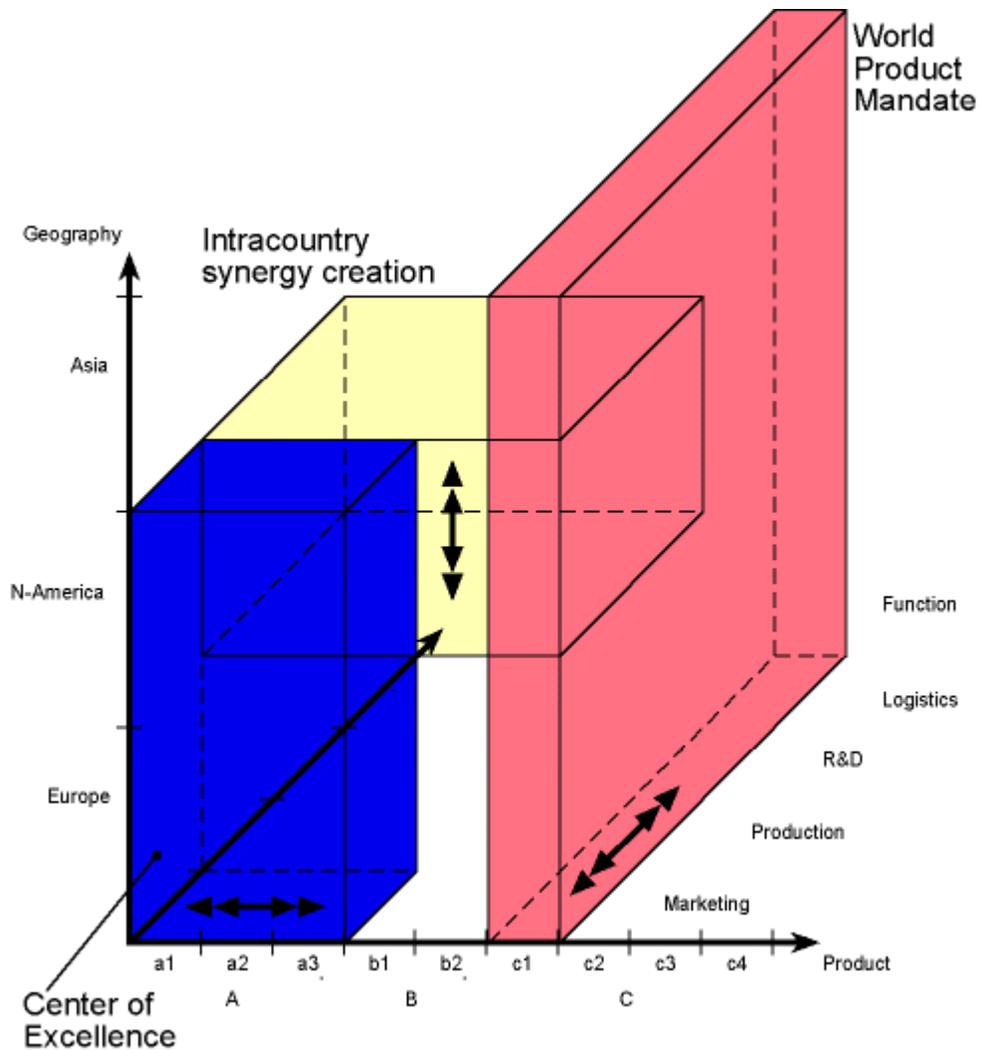

|          | course1 | course2 | course3 | course4 | course5 |
|----------|---------|---------|---------|---------|---------|
| student1 | 0 | 0 | 0 | 0 | 0 |
| student2 | 1 | 0 | 1 | 0 | 0 |
| student3 | 0 | 0 | 0 | 0 | 0 |
| student4 | 1 | 0 | 0 | 0 | 1 |
| student5 | 0 | 0 | 0 | 0 | 0 |
| student6 | 1 | 0 | 1 | 0 | 0 |

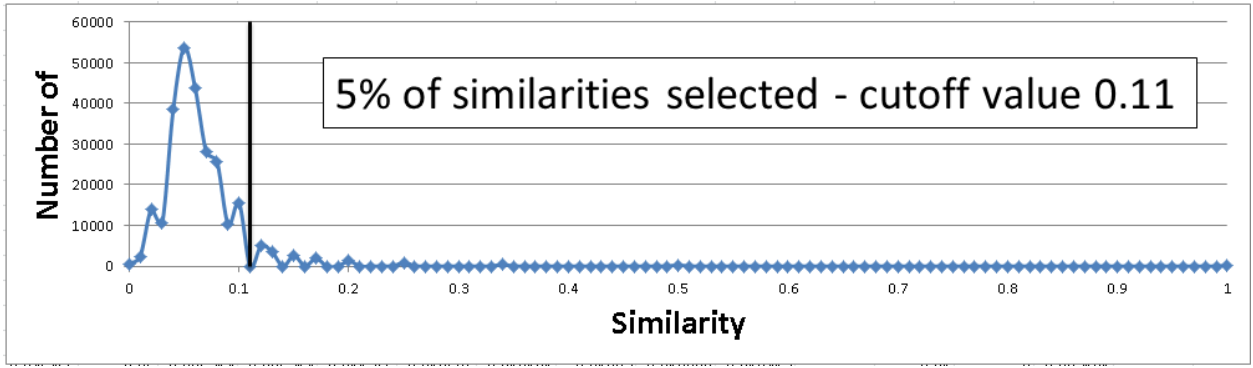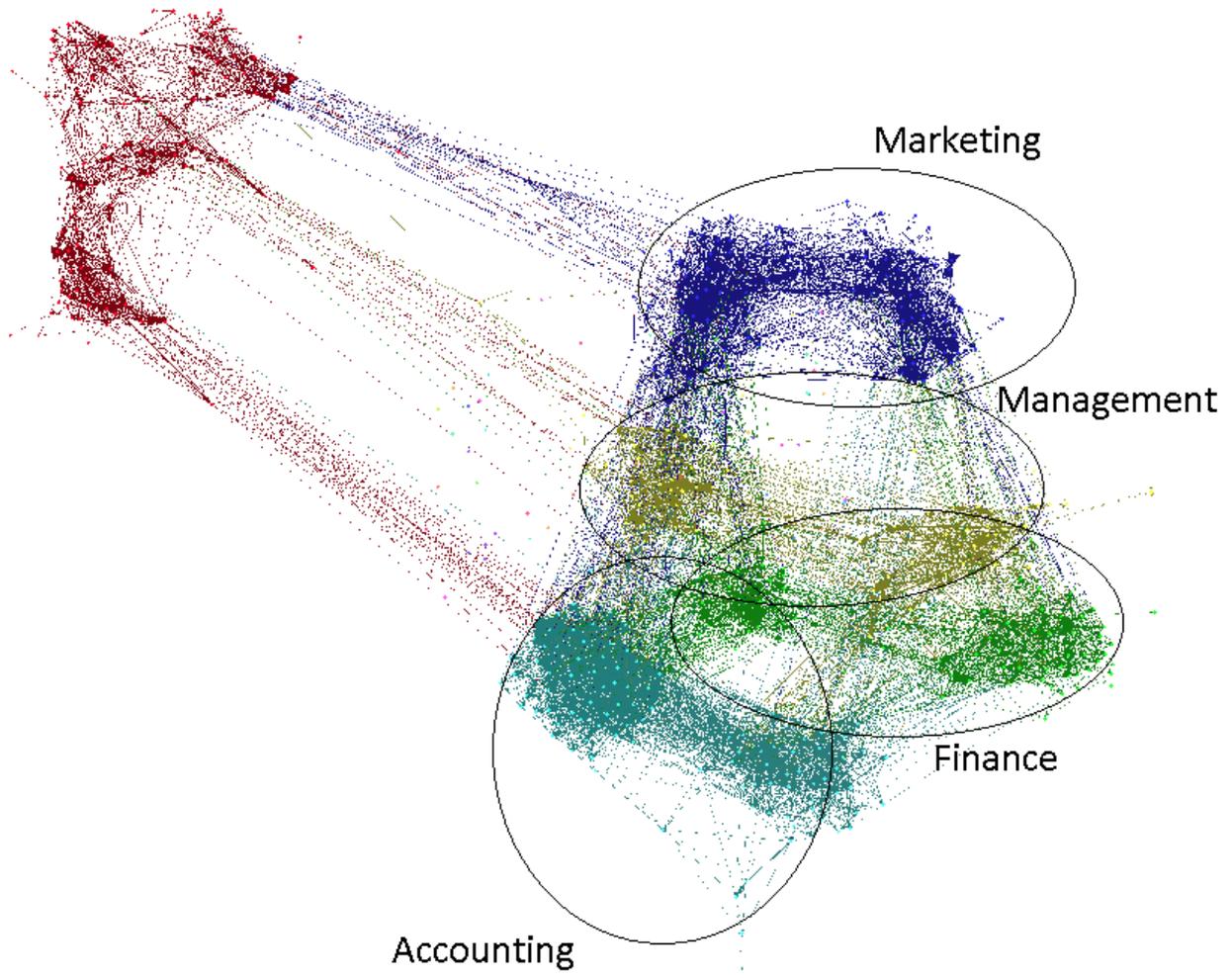

1. **The British Journal of Management has discussed the topic of coopetition (eg. Peng and Bourne, 2009; Ritala, 2011; Peng *et al.*, 2011). This paper follows the call by the same authors (Ritala, 2009) to investigate coopetition on levels different from the usual interfirm one. It discusses coopetition mainly on intrafirm level – internal coopetition. It also applies Eastern way of thinking, just as been done by Peng *et al.* (2011), in addressing the topic. The British Journal of Management has also actively discussed network science.**
2. **The reader will learn, that the correct way of thinking of internal coopetition, is as of a duality. Similarly to the relationship between light and dark, competition and cooperation are interdependent and defined by each other. The reader will learn that there very clearly exist coopetitive processes in multinational corporations and that processes taking place in universities are in parallel to these. Finally, the reader will learn, that these processes can be quantified by data mining, in order to assess the current state of affairs.**
3. **The fact that coopetition is pervasive, present on many different levels – societal, interfirm, intrafirm, personal, as well as elsewhere – is important. Even for an ordinary faculty member in a university, coopetition is an everyday reality. The fact that universities operate in ways that can be informed by the reality of multinational corporations is important for educational theorists. Data mining helps bring about the managerial implications of the research.**
4. **The social impact brought about is evident in terms of university management being on sounder grounds.**
5. **I hope that the paper will be cited for parallels drawn between MNC and university management, for the educational data mining component and also very importantly for furthering the coopetition discussion.**

# Data mining the MNC like internal co-opetition duality in a university context

Jaan Ubi1, Innar Liiv2, Evald Ubi3, Leo Vohandu4

Corresponding author: 1Deparment of Informatics, Tallinn University of Technology, Raja 15, Tallinn, 12618, Estonia, jaan.ubi@ttu.ee, cell: +372 56482063, fax: +372 6202305

2Deparment of Informatics, Tallinn University of Technology, Raja 15, Tallinn, 12618, Estonia, innar.liiv@ttu.ee

3Deparment of Economics, Tallinn University of Technology, Akadeemia tee 3, Tallinn, 12618, Estonia, evald.ubi@tseba.ttu.ee

4Deparment of Informatics, Tallinn University of Technology, Raja 15, Tallinn, 12618, Estonia, leo.vohandu@ttu.ee

Jaan Ubi is a Computer Science Ph.D. student in Tallinn University of Technology, where he has been employed as an Assistant Lecturer since 2004. His research interests include MNC corporations, educational data mining (biclustering and seriation) and operations research (linear planning and discriminant analysis). He is teaching spreadsheets, high level programming and optimization modeling to business and logistics students.

Innar Liiv received his Ph.D. in Computer Science in 2008 from Tallinn University of Technology, where he is currently an Associate Professor of Data Mining. His research interests include philosophy of artificial intelligence, data mining (seriation, clustering, association rules), predictive analytics (churn, fraud, lifetime value), operations research, business intelligence in logistics, information visualization and social network analysis.

Evald Ubi received his undergraduate degree in the Faculty of Mathematics in 1969 from Leningrad University and his Ph.D. in 1976 from Tallinn University of Technology. His research interests include linear planning and stochastic planning. He is an Associate Professor at the Department of Economics and teaches management science and operations research to business and logistics students.

Leo Vohandu received his Ph.D in Mathematics in 1955 from the University of Tartu. He is the founder of Tallinn University of Technology Institute of Infomatics (1966) and Professor Emeritus currently. He has supervised 44 Ph.D theses that have resulted in 16 current professorships. He is currently supervising 7 doctoral theses.